\documentclass[preprint,12pt]{elsarticle}

\usepackage{lineno,hyperref}
\usepackage{amsfonts}
\usepackage{amssymb}
\usepackage{amsthm,amsmath}
\usepackage{mathrsfs}
\usepackage{indentfirst}
\usepackage{multirow}

\journal{Physics Letters A}

\begin{document}

\begin{frontmatter}



\title{Defect-induced edge ferromagnetism and fractional spin excitations of the SU(4) $\pi$-flux Hubbard model on honeycomb lattice}


\author[inst1]{Yang Xie}
\author[inst1]{Cheng-Ping He}
\author[inst2]{Zhao-Yang Dong}
\ead{zhydong@njust.edu.cn}
\author[inst1,inst3]{Jian-Xin Li}
\ead{jxli@nju.edu.cn}

\affiliation[inst1]{organization={National Laboratory of Solid State Microstructures and Department of Physics,Nanjing University},
            addressline={22 Hankou Road},
            city={Nanjing},
            postcode={210093},
            state={Jiangsu},
            country={China}}

\affiliation[inst2]{organization={Department of Applied Physics, Nanjing University of Science and Technology},
            addressline={200 Xiaolingwei street},
            city={Nanjing},
            postcode={210094},
            state={Jiangsu},
            country={China}}

\affiliation[inst3]{organization={Collaborative Innovation Center of Advanced Microstructures, Nanjing University},
            addressline={22 Hankou Road},
            city={Nanjing},
            postcode={210093},
            state={Jiangsu},
            country={China}}

\begin{abstract}
\par Recently, a SU(4) $\pi$-flux Hubbard model on the honeycomb lattice has been proposed to study the spin-orbit excitations of $\alpha$-ZrCl$_3$ [Phys.~Rev.~Lett. 121.097201~(2017)]. Based on this model with a zigzag edge, we show the edge defects can induce edge flat bands that result in a SU(4) edge ferromagnetism. We develop an effective one-dimensional interaction Hamiltonian to study the corresponding SU(4) spin excitations. Remarkably, SU(4) spin excitations of the edge ferromagnet appear as a continuum covering the entire energy region rather than usual magnons. Through further entanglement entropy analysis, we suggest that the continuum consists of fractionalized spin excitations from the disappeared magnons, except for that from the particle-hole Stoner excitations. Moreover, in ribbon systems with finite widths, the disappeared magnons can be restored in the gap formed by the finite-size effect and the optical branch of the restored magnons are found to be topological nontrivial.
\end{abstract}



\end{frontmatter}


\section{Introduction}

\par Dirac Semimetals have been widely studied in recent years because of their unique electronic properties \cite{novoselov2005two, RevModPhys.81.109}. For example, there are edge flat-band states connecting the two Dirac points at zigzag boundaries in graphene \cite{Fujita1996, PhysRevB.54.17954, PhysRevB.59.8271, PhysRevB.77.085423}. Specially, the edge flat band will result the edge ferromagnetism when the Coulomb repulsion of electrons is taken into account \cite{PhysRevLett.99.177204, 2007First, PhysRevB.86.195418, PhysRevB.79.235433, PhysRevLett.106.226401, PhysRevB.86.115446, PhysRevLett.109.096404, matte2009novel, PhysRevB.81.245428, magda2014room, makarova2015edge,PhysRevB.102.224417}. Since the edge ferromagnetism is guaranteed by Tasaki's flat-band ferromagnetism \cite{PhysRevLett.69.1608, Hal1998From} and Lieb theory \cite{PhysRevLett.62.1201} for the SU(2) spin system, whether the edge ferromagnetism can be generalized into the SU(N) system is an open question. As we know, the SU(N) spin systems are usually realized in ultracold atomic systems, using the nuclear spin degrees of freedom \cite{Cazalilla_2014}. And it has been found the SU(N) symmetry can also be realized in electronic systems by combining spin and orbit degrees of freedom so that local electronic states are identified with a representation of SU(N) \cite{JPSJ.52.3897, JPSJ.66.1741, PhysRevX.2.041013, PhysRevB.91.155125, PhysRevLett.121.097201}. Recently, an effective SU(4) $\pi$-flux Hubbard model on the honeycomb lattice has been proposed for $\alpha$-ZrCl$_3$ due to the strong spin-orbit coupling \cite{PhysRevLett.121.097201}. Dirac points and edge states also exist in this system, so the edge flat bands and edge ferromagnetism are also expected in the system with zigzag boundaries.

\par In the paper, we show the edge defects can induce edge flat bands in the SU(4) $\pi$-flux honeycomb lattice model with zigzag boundaries that result in a SU(4) edge ferromagnetism once introducing the Hubbard interaction. To study the properties of the SU(4) edge ferromagnetism, we deveplop an effective Hamiltonian by projecting Hubbard interactions onto the edge flat band, and calculated the SU(4) spin excitations. Unlike the traditional ferromagnetic spin excitations, we find the excitations appear as a continuum covering the entire energy region with only magnon-like residuals. Through the further entanglement entropy (EE) analysis, we suggest that the residuals consist of fractionalized excitations of the disappeared magnons.
As a two-sublattice system, by analogy to the two branches of magnons, we can also find two parts of residuals in the continuum. Moreover, in ribbon systems with finite widths, the disappeared magnons can be restored in the gap formed by the finite-size effects and the optical branch of the restored magnons is shown to be topological nontrivial similar to that of Tasaki model \cite{PhysRevB.97.245111}.

\par The rest of the paper is organized as follows. In Sec. \ref{MM}, we firstly introduce the SU(4) $\pi$-flux model and its corresponding edge states with zigzag edges. Then we show the edge defect can induce edge flat bands. To study possible edge magnetism and corresponding spin excitations when the Hubbard interaction is included, the self-consistent mean-field approximation method and exact diagonalization method are introduced. In Sec. \ref{R}, we present the numerical results of the edge ferromagnetic ground state and its corresponding spin excitations. Section \ref{SD} provides a brief summary.

\section{Model and Method}\label{MM}

\subsection{SU(4) $\pi$-flux Hubbard model and edge states}

\begin{figure}
\centering
\includegraphics[scale=0.4]{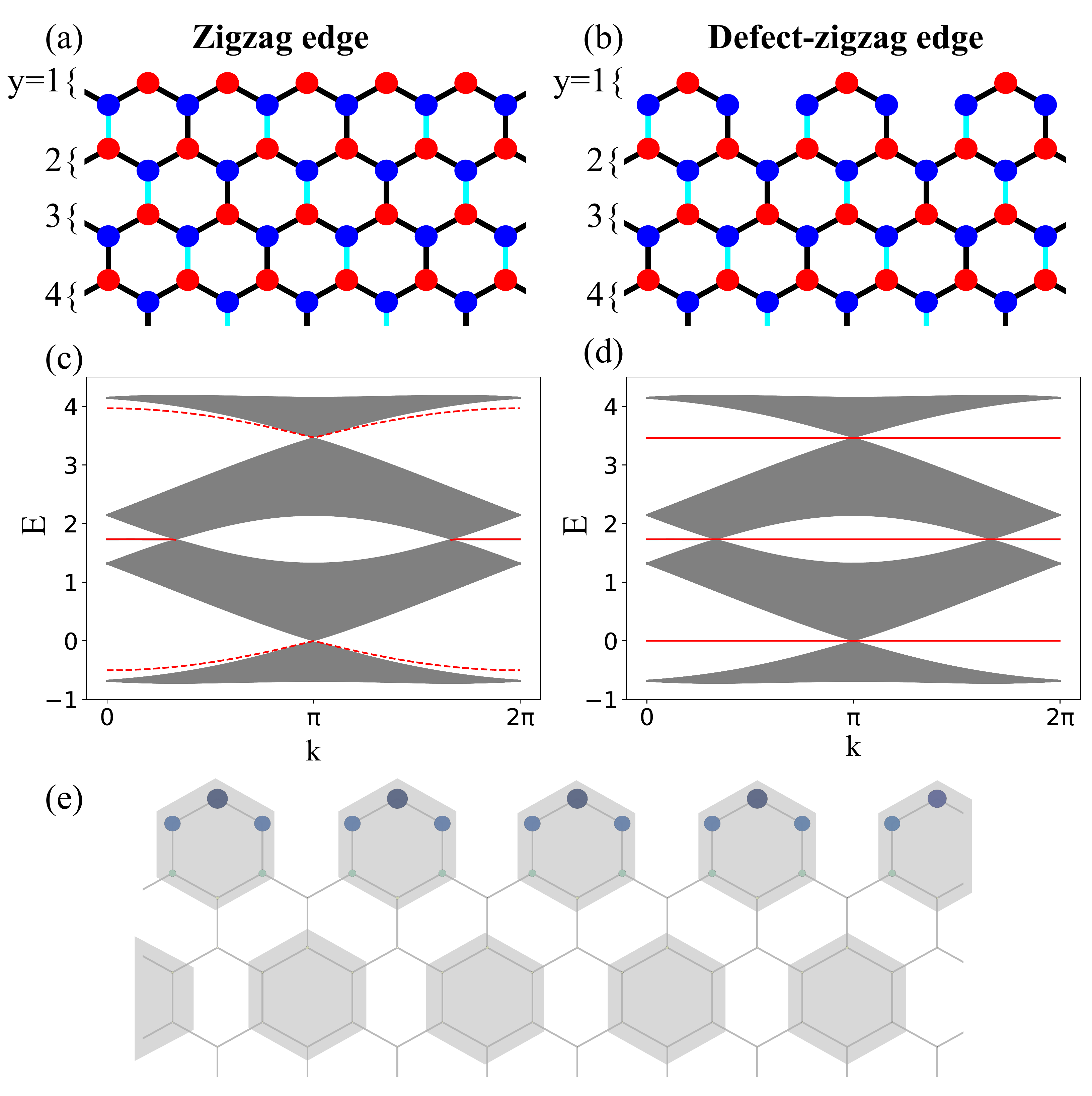}
\caption{Illustration of the SU(4) $\pi$-flux model in the honeycomb lattice with (a) zigzag edge, and (b) defect-zigzag edge, where the gauge $\varphi_{ij} = \pi$ for the cyan bonds and $\varphi_{ij} = 0$ for all others. (c) and (d) Energy band structure with the zigzag edge and defect-zigzag edge respectively, where the red lines denote the edge states. (e) Local density distribution of the flat band edge states denoted by the size of the filled circles. The gray hexagon indicates a unit cell of the single-particle real-space electronic wave functions.}\label{fig:Fig1}
\end{figure}

\par The model \cite{PhysRevLett.121.097201} we considered is written as $H=H_{0} + H_{U}$, where $H_{0}$ is the SU(4) $\pi$-flux model defined on the honeycomb lattice,
\begin{equation}
  H_{0}=-t\sum_{\langle i j\rangle \in \Lambda}\sum_{\sigma=1}^{4}(\mathrm{e}^{\mathrm{i} \varphi_{i j}} c_{i\sigma}^{\dagger} c_{j\sigma}+\text {H.c.}) + \mu \sum_{i \in \Lambda}\sum_{\sigma=1}^{4} c_{i\sigma}^{\dagger} c_{i\sigma},
\label{eq:Hamiltonian}
\end{equation}
and $H_{U}$ is the Hubbard interaction
\begin{equation}
  H_{U} =  U \sum_{i} \sum_{\sigma \neq \sigma^{\prime}} n_{i\sigma} n_{i \sigma^{\prime}}.
\end{equation}
Here, $\langle i j\rangle$ runs over all pairs of nearest-neighbor sites on lattice $\Lambda$, $\sigma=\sigma_1, \ldots, \sigma_4$ labels the flavor of a SU(4) fermion. The operator $c_{i\sigma}$ and its adjoint $c_{i\sigma}^{\dagger}$ satisfy the fermionic anticommutation relations $\left \{ c_{i \sigma}^{\dagger},c_{j \sigma^{\prime}} \right \}=\delta_{ij}\delta_{\sigma  \sigma^{\prime}}$ and $\left \{ c_{i  \sigma},c_{j \sigma^{\prime}} \right \}=0$ and the number operator for fermions at site $i$ is defined as $n_{i}=\sum_{\sigma =\sigma_1}^{\sigma_4} c_{i \sigma}^{\dagger}c_{i\sigma}$. $U$ is the on-site repulsive interaction. The hoping integral $t = 1$ is fixed as the unit and the chemical potential $\mu = \sqrt{3}$ for $1/4$ filling. We take a simple gauge $\varphi_{ij} = \pi$ for the cyan bonds (the rest $\varphi_{ij} = 0)$ as shown in Fig. \ref{fig:Fig1}~(a) to introduce the $\pi$-flux pattern. The $\pi$-flux creates new Dirac points at $k=\pi$ when the system is at $1/4$ filling and $3/4$ filling [Fig. \ref{fig:Fig1}~(c)]. so the edge flat bands are expected at zigzag boundaries.

\par Considering the zigzag boundary as shown in Fig. \ref{fig:Fig1}~(a), we take the Fourier transform of Eq.~\eqref{eq:Hamiltonian} along the $x$ direction,
\begin{equation}
  \mathbf{H_{0}}=\sum_{k \sigma} \mathbf{C}_{k \sigma}^{\dagger}\left(\begin{array}{cccc}
  \mathbf{A}(k) & \mathbf{B}(k) & \cdots & 0 \\
  \mathbf{B^{\dagger}}(k) & \mathbf{A}(k) & \cdots & 0 \\
  \vdots & \vdots & \ddots & \vdots \\
  0 & 0 & \ldots & \mathbf{A}(k)
\end{array}\right) \mathbf{C}_{k \sigma},
\label{Hamiltonian_semi}
\end{equation}
where $\mathbf{C}_{k \sigma}^{\dagger}=(c_{1 k \sigma}^{\dagger}, c_{2 k \sigma}^{\dagger}, \cdots)$ is the creation operators along the $y$ direction, and $\mathbf{A}(k)$, $\mathbf{B}(k)$ read,
\begin{equation}
  \mathbf{A}(k)=\left(\begin{array}{cccc}
  \sqrt{3}t & 0 & t & t\mathrm{e}^{-\mathrm{i} k} \\
  0 & \sqrt{3}t & t & t \\
  t & t & \sqrt{3}t & 0 \\
  t\mathrm{e}^{\mathrm{i} k} & t & 0 & \sqrt{3}t
\end{array}\right),
~\mathbf{B}(k)=\left(\begin{array}{cccc}
  0 & 0 & 0 & 0 \\
  0 & 0 & 0 & 0 \\
  t & 0 & 0 & 0 \\
  0 & -t & 0 & 0
\end{array}\right).
\end{equation}
 We diagonalize Eq. \eqref{Hamiltonian_semi} to obtain the energy band of the free part of the Hamiltonian, and the result is presented in Fig. \ref{fig:Fig1}~(c). One can find that the edge states at 1/4 and 3/4 fillings are not flat as denoted by the red dashed lines. In order to flatten the edge states, we introduce defects at the edge as shown in Fig. \ref{fig:Fig1}~(b). In other words, the edge states can be localized to be a flat band by the defects. The result of the energy band with the defected zigzag edge is shown in Fig. \ref{fig:Fig1}~(d). Now, we find that three flat bands across the whole Brillouin zone (BZ) appear, the upper and lower flat bands are particle-hole symmetric, and the middle one connects the original one. Here, we will focus on the lower flat band at zero energy with $1/4$ filling. The quasiparticles of this band read $
d_{k \sigma}=\boldsymbol{\mu}_{k} \mathbf{C}_{k \sigma}$,
where $\boldsymbol{\mu}_{k} = \left(\boldsymbol{\nu}_{k}, \lambda \boldsymbol{\nu}_{k}, \cdots, \lambda^{i-1} \boldsymbol{\nu}_{\boldsymbol{k}}, \cdots\right)$ is the corresponding eigenvector of parameter matrix of Eq.~\eqref{Hamiltonian_semi} and the expression of $\boldsymbol{\nu_{k}}$ is given by
\begin{equation}
\begin{aligned}
\boldsymbol{\nu_{k}}=&\left(0, \frac{2+2 \mathrm{e}^{-\mathrm{i} k}+s_{k}}{\sqrt{3}}, \frac{-3-\mathrm{e}^{-\mathrm{i} k}-s_{k}}{2},\right.\\
& \frac{-1-3 \mathrm{e}^{-\mathrm{i} k}-s_{k}}{2}, \frac{5-\mathrm{e}^{-\mathrm{i} k}+s_{k}}{2 \sqrt{3}}, \frac{1-5 \mathrm{e}^{-\mathrm{i} k}-s_{k}}{2 \sqrt{3}}, \\
&\left.\mathrm{e}^{-\mathrm{i} k}-1,0\right), \\
\lambda=& \frac{-1-\mathrm{e}^{-2 \mathrm{i} k}-6 \mathrm{e}^{-\mathrm{i} k}+\left(1+\mathrm{e}^{-\mathrm{i} k}\right) s_{k}}{2 \mathrm{e}^{-\mathrm{i} k}-2},
\end{aligned}
\end{equation}
where $\lambda$ is the decay factor satisfying $|\lambda| < 1$ and $s_{k}=\sqrt{1+14 \mathrm{e}^{-\mathrm{i} k}+\mathrm{e}^{-2 \mathrm{i} k}}$. The local density distribution of
the flat band edge states are shown by the different sizes of the filled circles in Fig. \ref{fig:Fig1}~(e), it shows a rapid decay into the bulk.

\subsection{Self-consistent mean-field approximation}

\par According to the generalization of the flat-band ferromagnetism theory to the SU(N) Hubbard model \cite{PhysRevLett.69.1608, Hal1998From, PhysRevB.100.214423, LIU20191490}, an edge ferromagnetism will emerge here. To verify the edge SU(4) ferromagnetic order, we use the
self-consistent mean-field approximation for the SU(4) $\pi$-flux  Hubbard model \cite{PhysRevA.96.053616}. The mean-field Hamiltonian can be written as follows,
\begin{equation}
  H_{\mathbf{MF}}=H_{0}-\frac{16}{5} U \sum_{i, \sigma \sigma'} c_{i \sigma}^{\dagger} \mathbf{m}_{i} \cdot \frac{\mathbf{\Gamma}_{\sigma \sigma'}}{2} c_{i \sigma'}+\frac{8}{5} U \sum_{i} \mathbf{m}_{i}^{2},
\label{eq:SU4MeanField}
\end{equation}
where $H_{0}$ is given by Eq. \eqref{eq:Hamiltonian}, and the SU(4) spin order parameters (magnetic moments) $\mathbf{m}_{i}=\langle \mathbf{S}_{i}\rangle$ are given by,
\begin{equation}
  \hat{S}_{i}^{a}=\sum_{\sigma, \sigma'=1}^{4} c_{i \sigma}^{\dagger} \Gamma_{\sigma \sigma'}^{a} c_{i \sigma'} \quad(a=1, \ldots, 15),
\label{eq:SU4Spin}
\end{equation}
where $\Gamma_{\sigma \sigma'}^{a}$ are the generators of the SU(4) Lie algebra which are the matrix elements of Dirac matrices \cite{hladik1999spinors, PhysRevB.97.205106}. For the general SU(4) magnetic order, 15 order parameters $\mathbf{m}_{i}$  are required to describe the magnetic properties of the system. Since the magnetic moment vectors on all sublattices of a collinear magnetic state are parallel or antiparallel to each other, a global SU(4) unitary transformation on Eq. \eqref{eq:SU4MeanField} can be performed to eliminate redundant degrees of freedom \cite{PhysRevA.96.053616}.
\begin{equation}
\begin{aligned}
  H^{\prime}_{\mathbf{MF}}=& \mathbf{U}H_{\mathbf{MF}}\mathbf{U}^{\dagger}\\
  =& H_{0}-\frac{16}{5} U \sum_{i, \sigma \sigma'} c_{i \sigma}^{\dagger} \widetilde{\mathbf{m} }_{i} \cdot \frac{\widetilde{\mathbf{\Gamma}}_{\sigma \sigma'}}{2} c_{i \sigma'}+\frac{8}{5} U \sum_{i} \widetilde{\mathbf{m} }_{i}\cdot  \widetilde{\mathbf{m} }_{i},
\label{eq:SU4MeanField2}
\end{aligned}
\end{equation}
where $\widetilde{\mathbf{\Gamma}}$ are the three diagonal commuting generators in the SU(4) algebra and constitute the Cartan sub-algebra of the SU(4) Lie algebra \cite{georgi2000lie}. The SU(4) collinear magnetic orders can be described by these three diagonal order parameters $\widetilde{m}^{a}=\langle \widetilde{\Gamma}^a\rangle$, whose operators are commutative with each other and $\mathbf{S}_{i}^2$, just like $S^z$ for the SU(2) case. The three order parameters are given by,
\begin{equation}
\begin{aligned}
  &\widetilde{m}_{i}^{1}=\left(\left\langle n_{i \sigma_{1}}\right\rangle-\left\langle n_{i \sigma_{2}}\right\rangle\right) / 2, \\
  &\widetilde{m}_{i}^{2}=\left(\left\langle n_{i \sigma_{1}}\right\rangle+\left\langle n_{i \sigma_{2}}\right\rangle-2\left\langle n_{i \sigma_{3}}\right\rangle\right) / 2 \sqrt{3}, \\
  &\widetilde{m}_{i}^{3}=\left(\left\langle n_{i \sigma_{1}}\right\rangle+\left\langle n_{i \sigma_{2}}\right\rangle+\left\langle n_{i \sigma_{3}}\right\rangle-3\left\langle n_{i \sigma_{4}}\right\rangle\right) / 2 \sqrt{6},
\end{aligned}
\label{eq:SU4Order}
\end{equation}
where $n_{i\sigma} = c_{i \sigma}^{\dagger}c_{i \sigma}$ is the particle number operator at site $i$ with flavor $\sigma$ and $\sigma_1, \sigma_2, \sigma_3, \sigma_4$ indicate the four flavors. The Eqs. \eqref{eq:SU4MeanField2} and \eqref{eq:SU4Order} can be solved by the self-consistent numerical calculations.

\subsection{Calculations of spin excitations}


\par Due to the edge flat bands are dominated by the Hubbard interaction, we can project the Hubbard interactions onto the flat band to study the excitations \cite{PhysRevB.102.224417, PhysRevB.97.245111, PhysRevB.99.014407, PhysRevB.85.075438, PhysRevLett.106.236804, PhysRevX.1.021014, PhysRevB.84.165107, PhysRevB.83.195432, PhysRevLett.108.046806},
\begin{equation}
\begin{aligned}
  H_{\mathrm{eff}}=& P \hat{H} P \\
  =& \sum_{k \sigma} \varepsilon(k) d_{k \sigma}^{\dagger} d_{k \sigma} \\
  &+\frac{U}{2 N} \sum_{y \sigma \sigma^{\prime}} \sum_{k k^{\prime} p} \mu_{y k+p \sigma}^{*} \mu_{y k^{\prime}-p \sigma^{\prime}}^{*} \mu_{y k^{\prime} \sigma^{\prime}} \mu_{y k \sigma}
    d_{k+p \sigma}^{\dagger} d_{k^{\prime}-p \sigma^{\prime}}^{\dagger} d_{k^{\prime} \sigma^{\prime}} d_{k \sigma},
\end{aligned}
\label{Hamiltonian_eff}
\end{equation}
where $P$ is the projection operator, $\mu_{y k \sigma}$ are the elements of the eigenvectors of Eq.~\eqref{Hamiltonian_semi} mentioned above, and $\varepsilon(k)$ is the dispersion of the flat band which is assumed to be zero for simplicity. It is known that the ferromagnetic ground state of the edge ferromagnetism is suggested to be the state with all the same flavor $\sigma_1$, written as $|\mathrm{FM}\rangle \equiv  \prod_{k \in \mathrm{FBZ}} d_{k \sigma_1}^{\dagger}|\Omega\rangle$ \cite{PhysRevLett.69.1608,Hal1998From, PhysRevB.100.214423, LIU20191490}, where $d_{k \sigma_1}^{\dagger}$ is the quasiparticle creation operator of flat-band edge states of flavor $\sigma_1$ and $|\Omega\rangle$ is the ground state of bulk. Thus the single-flavor excitations from the ground state flavor $\sigma_1$ to another flavor $\sigma$ can be written as $\left|k_{i}\right\rangle_{q}=d_{k_{i} \sigma}^{\dagger} d_{k_{i}-q \sigma_1}|\mathrm{FM}\rangle$. In this base, we can calculate the matrix element of Eq.~\eqref{Hamiltonian_eff} and the results are,
\begin{equation}
  \left\langle k_{j}\left|H_{\mathrm{eff}}\right| k_{i}\right\rangle_{q}=M_{i}^{s}(q) \delta_{k_{j}, k_{i}}-M_{j i}^{m}(q)
  \label{eq:EigenEquation},
\end{equation}
in which,
\begin{equation}
\begin{aligned}
  M_{i}^{s}(q) &=\frac{U}{N} \sum_{y}\sum_{p}\left|\mu_{y p \sigma_1}\right|^{2}\left|\mu_{y k_{i}-q \sigma}\right|^{2},
\end{aligned}
\label{eq:StonerTerm}
\end{equation}
\begin{equation}
\begin{aligned}
  M_{j i}^{m}(q) &=\frac{U}{N} \sum_{y}  \mu_{y k_{i}-q \sigma}^{*} \mu_{y k_{i} \sigma_1} \mu_{y k_{j}-q \sigma} \mu_{y k_{j} \sigma_1}^{*}.
\end{aligned}
\label{eq:MagnonTerm}
\end{equation}
Considering that the amplitude $\mu_{yk\sigma}$ exhibits an exponential decay when entering into the bulk, we limit the summation of $y$ here only to the edge unit cell. It is known that two main kinds of spin excitations of the itinerant ferromagnet are Stoner continuum and magnons. The solutions of diagonal matrix Eq.~\eqref{eq:StonerTerm} are the individual excitations making up the Stoner continuum, and the nontrivial solutions of Eq.~\eqref{eq:MagnonTerm} determine the bases for magnons coupled to the individual excitations.

With the projected Hamiltonian Eq.~\eqref{eq:EigenEquation}, we can calculate the single-flavor excitations of the SU(4) edge ferromagnet using the exact diagonalization method, and their correlation function is then calculated by,
\begin{equation}
  S^{\sigma \sigma^{\prime}}(q, \omega)=\frac{1}{2 \pi} \int \sum_{y, y^{\prime}}\left\langle\tilde{S}_{y}^{\sigma \sigma^{\prime}}(-q, 0) \tilde{S}_{y^{\prime}}^{\sigma^{\prime} \sigma}(q, t)\right\rangle \mathrm{e}^{\mathrm{i} \omega t} dt,
\label{eq:SpinCorrelation}
\end{equation}
where $\tilde{S}^{\sigma\sigma^{\prime}}_{y}(q)=\sum_{k}\mu^{*}_{k\sigma}\mu_{k-q\sigma^{\prime}}d_{k \sigma}^{\dagger} d_{k-q\sigma^{\prime}}$. At the meantime, we can calculate the spectral function of the single-flavor excitations after obtaining the eigenvalue $E_{i}(q)$  and corresponding eigenvector $|\Psi_{i}(q)\rangle$ of Eq.~\eqref{eq:EigenEquation}, which is written by,
\begin{equation}
  A^{\sigma \sigma^{\prime}}(q, \omega)=-\frac{1}{\pi} \operatorname{Im}\left[\sum_{i} \frac{\left|\left\langle P^{\sigma \sigma^{\prime}}(q) \mid \Psi_{i}(q)\right\rangle\right|^{2}}{\omega-E_{i}(q)+\mathrm{i} \eta}\right],
\end{equation}
in which $\left|P^{\sigma \sigma^{\prime}}(q)\right\rangle=\tilde{S}_{y}^{\sigma \sigma^{\prime}}(q)|\mathrm{FM}\rangle$. In the following analysis, to further check the excitation spectra excluding the individual Stoner excitations, we define the collective excitation spectra by using the nontrivial solutions $\left|\psi_{m}(q)\right\rangle$ of Eq.~\eqref{eq:MagnonTerm},
\begin{equation}
  A^{m}(q, \omega)=-\frac{1}{\pi} \operatorname{Im}\left[\sum_{i} \frac{\left|\left\langle\psi_{m}(q) \mid \Psi_{i}(q)\right\rangle\right|^{2}}{\omega-E_{i}(q)+\mathrm{i} \eta}\right],
\label{eq:magnonSpectra}
\end{equation}
where $\eta$ is taken to be $0.005$ in the calculations.

\subsection{Entanglement entropy analysis}
\label{EE}
\par The EE analysis method is used to characterize the spin excitation modes for this system \cite{PhysRevB.102.224417, PhysRevB.104.L180406}. The eigenstates of Eq.~\eqref{eq:EigenEquation} can be decomposed into the direct product of two parts
\begin{equation}
  \left|\Psi_{i}(q)\right\rangle=\sum_{k} \psi_{i}(k)\left|\sigma_{k-q}\right\rangle \otimes\left|\bar{\sigma_1}_{k}\right\rangle,
  \label{eq:Schmidt}
\end{equation}
where $\left|\sigma_{k-q}\right\rangle$ and $\left|\bar{\sigma_1}_{k}\right\rangle$ represent a particle in flavor $\sigma$ space and a hole in the flavor $\sigma_1$ space. Eq.~\eqref{eq:Schmidt} is also the Schmidt decomposition of particle-hole pair. So the EE of $\left|\Psi_{i}(q)\right\rangle$ can be defined as follows with respect to this bipartition
\begin{equation}
S_{i}(q)=-\sum_{k}\left|\psi_{i}(k)\right|^{2} \ln \left|\psi_{i}(k)\right|^{2}.
\label{eq:EE}
\end{equation}
Different excitations can be identified by the scaling behavior of EEs \cite{PhysRevB.102.224417}. The EEs of free particle-hole pairs will converge to a constant, while the EEs of the magnons formed by the confined particle-hole pairs will diverge logarithmically with the system size.

\section{Results}\label{R}

\subsection{Edge ferromagnetism}

\begin{figure}
\centering
\includegraphics[scale=0.4]{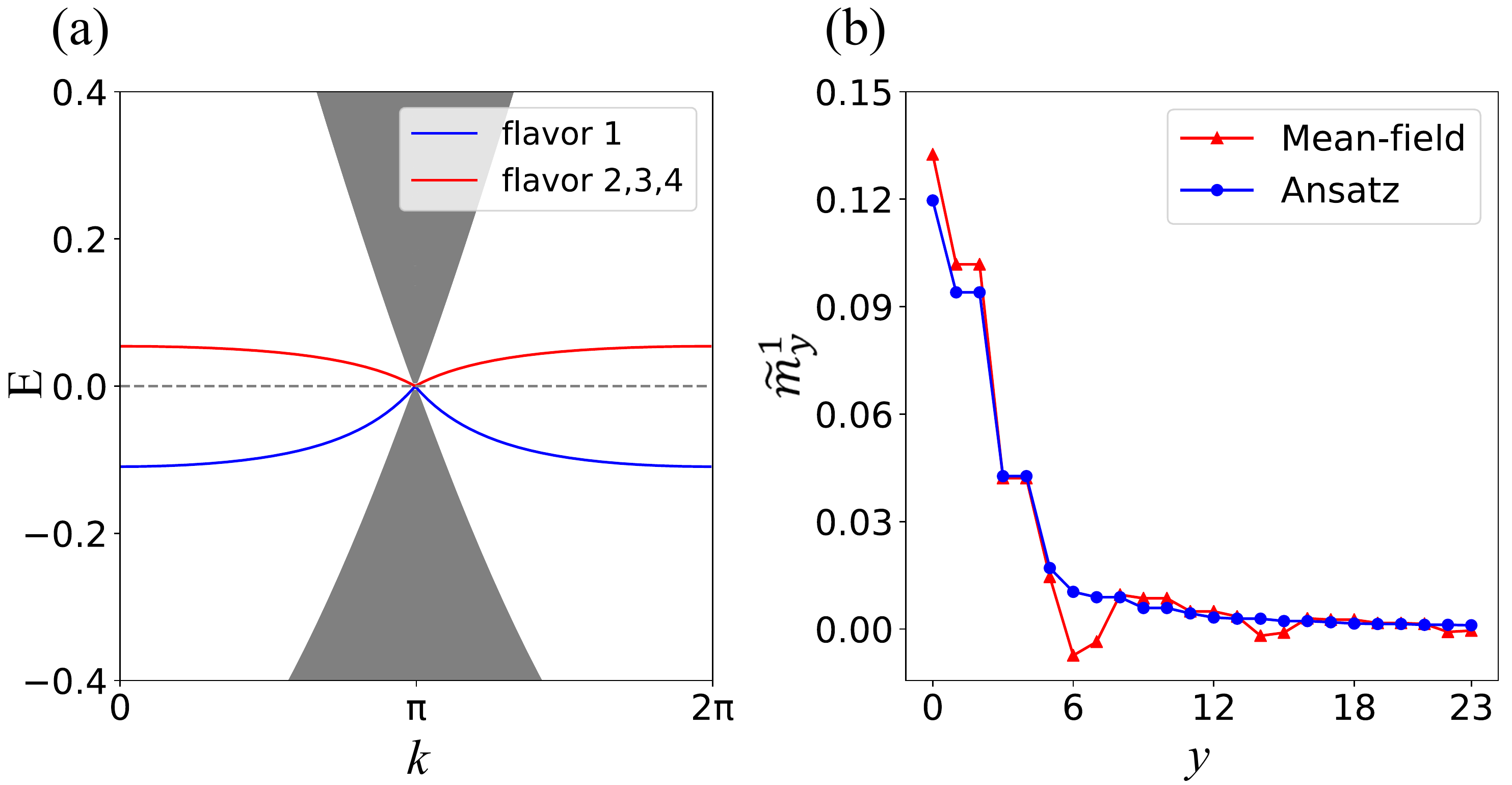}
\caption{(a) Energy band structure and (b) magnetic moment distribution of the half-infinite SU(4) $\pi$-flux Hubbard model with defect-zigzag edges calculated with the mean-field approximation and $U=0.5t$.} \label{fig:Fig2}
\end{figure}

\par We firstly calculate the energy band and magnetic moment distribution of the half-infinite SU(4) $\pi$-flux Hubbard model with the defect-zigzag edge using the mean-field approximation where we take $U=0.5t$. The energy band is shown in Fig. \ref{fig:Fig2}~(a), we find that the four-fold degenerate flat band is broken into a non-degenerate band denoted by the read line and a three-fold degenerate band by the blue line with the introduction of the Hubbard interaction $U$. For a 1/4 filling discussed in this paper, the non-degenerate band is below the Fermi level with
$E=0$, which is filled with one species of SU(4) spins, such as the flavor $\sigma_1$ shown in Fig. \ref{fig:Fig2}~(a).
Due to the degeneracy of other three edge states, only one order parameter $\widetilde{m}_{y}^1 \equiv (n_{y\sigma_1} - n_{y\sigma}) / 2$ is needed to describe the system similar to the case in the SU(2) spin system, where $\sigma_1$ ($\sigma$) represents the flavor of the occupied (unoccupied) edge states. The calculated magnetization distribution is shown in Fig. \ref{fig:Fig2}~(b) denoted by the red line. It shows a finite magnetic moment, but the magnetic moments are localized near the edge and decays nearly exponentially towards the bulk, which shares a similarity with the distribution of the edge state LDOS in Fig. \ref{fig:Fig1}~(d). These results suggest that the SU(4) edge ferromagnetic state exists in the SU(4) $\pi$-flux Hubbard model with a defect-zigzag edge.

According to the generalization of the flat-band ferromagnetism theory to the SU(N) Hubbard model \cite{PhysRevLett.69.1608, Hal1998From, PhysRevB.100.214423, LIU20191490}, the ground state is suggested to be $|\mathrm{FM}\rangle \equiv  \prod_{k \in \mathrm{FBZ}} d_{k \sigma_1}^{\dagger}|\Omega\rangle$ as described above. Based on this ansatz, the magnetic moment of the edge ferromagnetism state can also been calculated by $\widetilde{m}_{y}^1\equiv \sum_{k}(\mu^{*}_{k\sigma_1}\mu_{k\sigma_1}d_{k \sigma_1}^{\dagger}d_{k\sigma_1}-\mu^{*}_{k\sigma}\mu_{k\sigma}d_{k \sigma}^{\dagger} d_{k\sigma})/2$. The results are presented in Fig. \ref{fig:Fig2}~(b) as denoted by the blue line. One can find that the distribution of the magnetic moment calculated with the ansatz basically coincides with the mean-field result. This coincidence gives a strong support on our suggestion of the SU(4) edge ferromagnetic state.
Additionally, we also find other two SU(4) ferromagnetic states when the edge states are $1/2$ filled and $3/4$ filled, where the edge states with the $3/4$ filling and $1/4$ filling in Fig.~\ref{fig:Fig2} can be related by the particle-hole symmetry \cite{SM}.


\subsection{Spin excitations}\label{Paradox}

\begin{figure}
\centering
\includegraphics[scale=0.4]{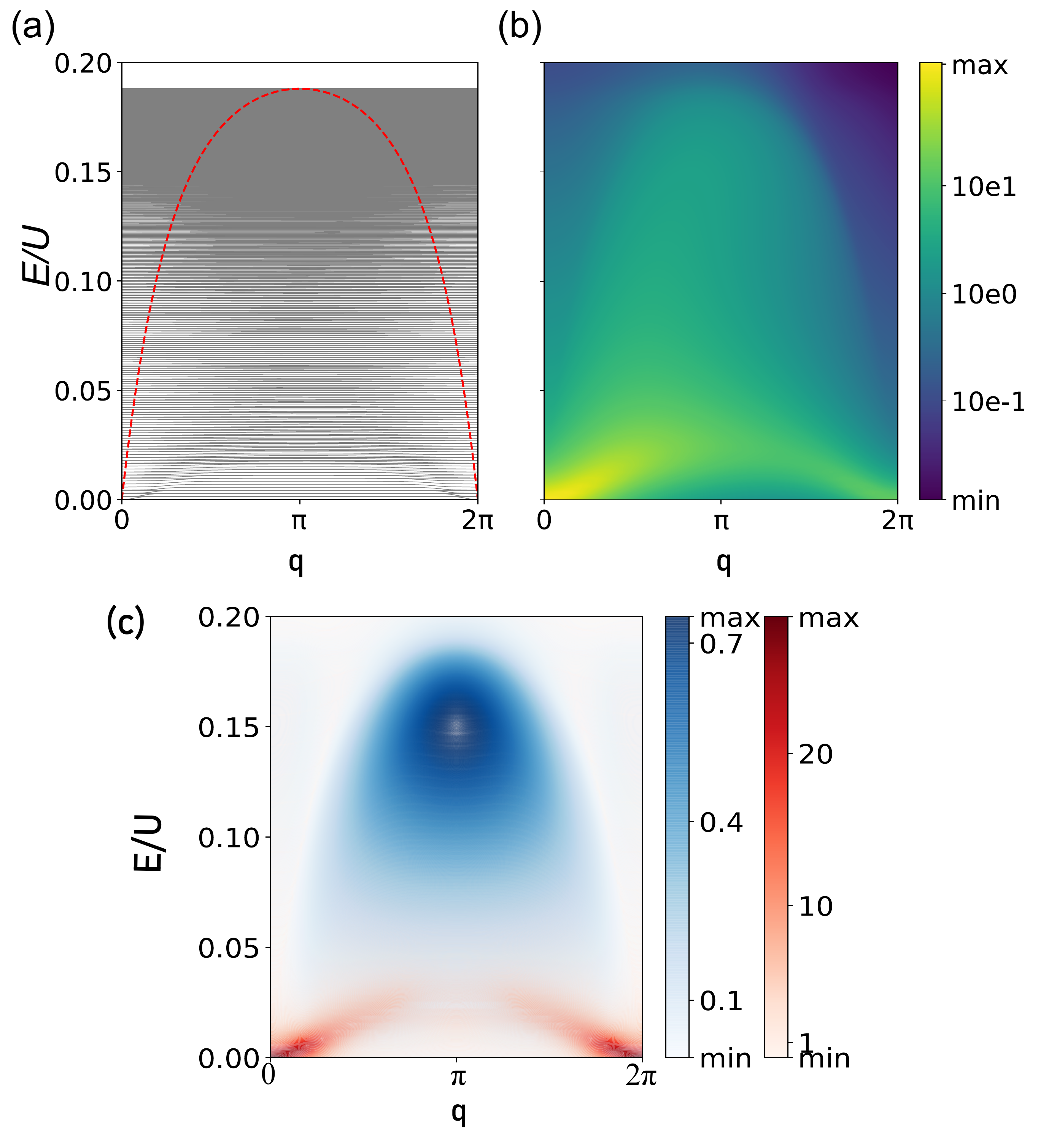}
\caption{(a) Dispersion and (b) spectral functions of single-flavor excitations and (c) single magnon spectra calculated by Eq.~\eqref{eq:magnonSpectra} over the edge ferromagnetic ground state of half-infinite SU(4) $\pi$-flux model with defect-zigzag edges. The red and the blue denote the spectra from the two major branches of magnonic solutions of Eq.~\eqref{eq:MagnonTerm}.}\label{fig:Fig3}
\end{figure}

\par The spin excitation dispersion and spectral function are shown in Fig.~\ref{fig:Fig3}~(a) and (b), respectively. As is well known, for an itinerant magnetic state such as the flat-band ferromagnet discussed here, there are high-energy Stoner continuum and low-energy magnons. The feature of the SU(4) spin excitations shown in Fig.~\ref{fig:Fig3}~(a) and (b) is that the continuum covers the entire energy region and there is a clear upper boundary in Fig.~\ref{fig:Fig3} (b) which is denoted by the red line in Fig.~\ref{fig:Fig3}~(a). The boundary corresponds to the eigenmode $d_{\pi-q \sigma}^{\dagger} d_{\pi \sigma_1}|\mathrm{FM}\rangle$. 
We can identify the dispersionless excitations beyond the boundary as the Stoner continuum, as these particle-hole excitations come from the flat band.
From Fig.~\ref{fig:Fig3}~(a), we find that the Stoner continuum reaches down to zero energy, so that it will interact with the collective modes. As a result, the whole spectra exhibits a continuum.
To see whether magnons can be distinguished as collective modes, superimposed on the continuum, we should check the features of Fig. \ref{fig:Fig3}~(a) and (b) in more detail. One will find that the nearly flat dispersions at low energies in Fig. \ref{fig:Fig3}~(a) are distorted and the distortions appear to form a reminiscence of a ferromagnetic magnonic dispersion in the continuum starting parabolically from $q=0, 2\pi$ and reach the highest energy at $q=\pi$. The reminiscence is related to the dominate modes with the high spectral weights in the excitation spectra in Fig. \ref{fig:Fig3}~(b). However, from the spectral weights we expect that these are not well defined magnons due to their broad spectral broadening when immersed in the continuum, and we term them as dominate modes hereafter. To better observe this, we check the excitation spectra by excluding the individual Stoner excitations which is determined by Eq.~\eqref{eq:StonerTerm}. To do this, we use Eq.~\eqref{eq:magnonSpectra} to calculate the spectra coming from the collective excitations and the results are presented in Fig.~\ref{fig:Fig3}~(c). In this case, the spectra evolved from the two branches of the collective excitations can be clearly distinguished. 
The lower branch seems to be heavily dampened and not well-defined magnons, and more excessively, the upper one losses its dispersion and concentrates at the centre of the continuum around $q=\pi$.

\begin{figure}
\centering
\includegraphics[scale=0.4]{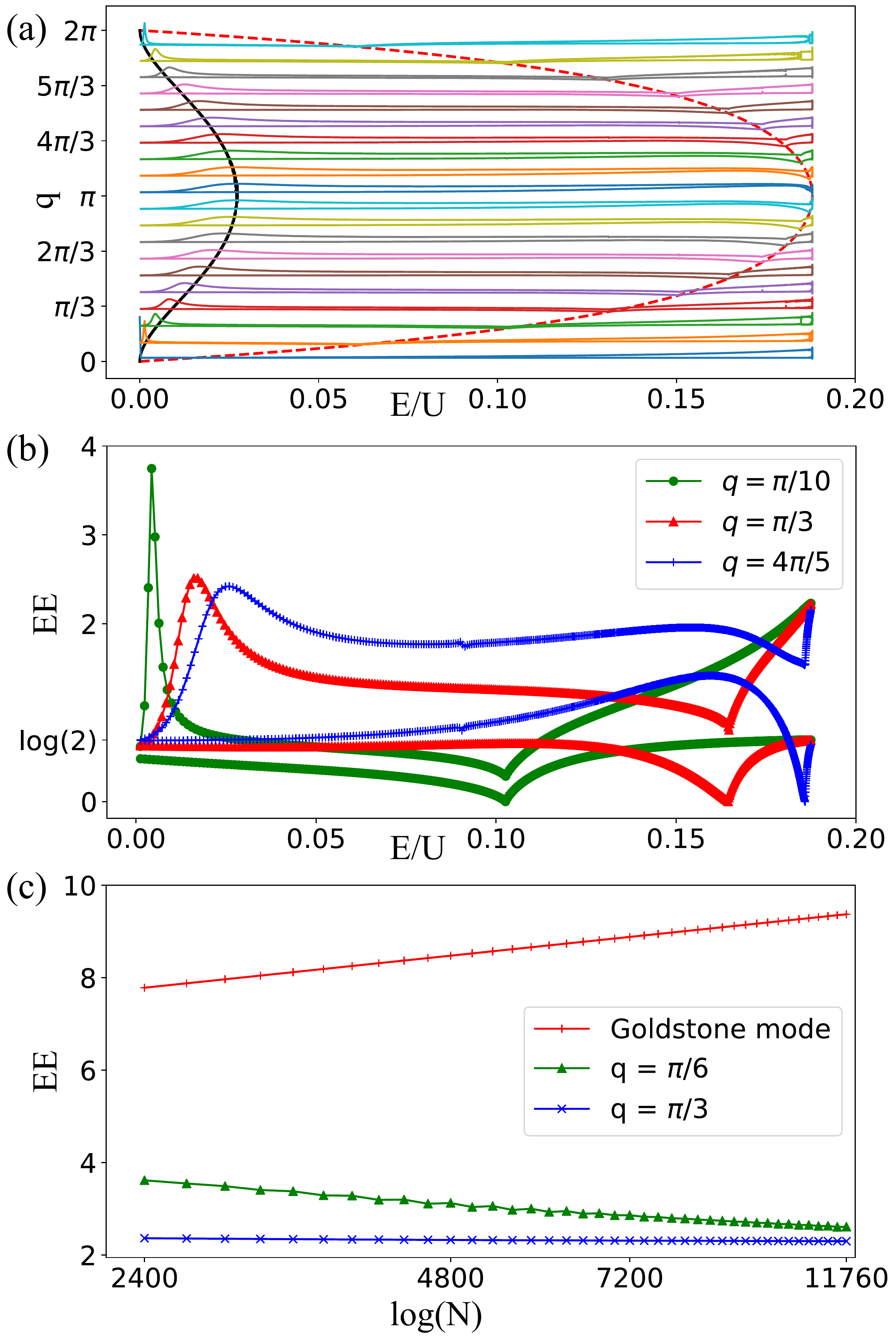}
\caption{(a) EE spectra of single-flavor excitations. The black line indicates the  maxima of the EEs and the red line denotes the boundary. (b) EEs at q = $\pi/10$ (green), q = $\pi/3$ (red), and q = $4\pi/5$ (blue). (c) EE scaling behaviors of the maximum EE modes at q = $\pi/6$ (green), q = $\pi/3$ (blue) and the Goldstone mode q = 0 (red).}\label{fig:Fig4}
\end{figure}

\par To further quantify the nature of these dominate modes, we will resort to the EEs analysis introduced in Sec.~\ref{EE}. Let us first discuss the general features of the EE spectra as shown in Fig. \ref{fig:Fig4}~(a). We can also find that two branches of the EE spectra exists for each momentum $q$, this can seen more clearly from Fig.~\ref{fig:Fig4}~(b) where only three typical EE spectra for $q=\pi/10, \pi/3, 4\pi/5$ are presented. Among them, one branch has a peak at a low energy for each $q$, which form a dispersion denoted by the black line. The peak position of EEs is right at the dominate mode energy and its maximum spectral weight shown in Fig.~\ref{fig:Fig3}~(b). Moreover, the height of the EE peak decreases with the increase of $q$ from $0$ to $\pi$, which is also consistent with the distribution of the spectral weight of the dominate mode. So, we can associate the peak in the EE spectra with the dominate mode in the spin excitation spectra. Another branch of the EE spectra has a lower EE, and distributes around $\ln 2$ in the low energy region and converges to $\ln 2$ at the top of the spin excitation spectrum. On the other hand, the EE for $q$ around $\pi$, such as $q=4\pi/5$ denoted by the blue line in Fig.~\ref{fig:Fig4}~(b), increases in the high energy region. This is because the additional contributions from the other branch of the collective excitations centered around $q=\pi$ as has been discussed based on Fig.~\ref{fig:Fig3}~(c). At the boundary of the particle-hole individual modes denoted by the dashed red line, the EEs exhibit a clear dip. Therefore, we find that the EEs can been used to characterize the spin excitations.

\par Then, we study the scaling behaviors of the EEs with the system size $N$. The EEs of the dominate mode calculated using the peak value in the EE spectra as a function of $N$ for two different $q$ with $q=\pi/6$, $q=\pi/3$ are presented Fig.~\ref{fig:Fig4}~(c), together with that for $q=0$ representing the Goldstone mode.
We see that the EE for the Goldstone mode is logarithmically divergent with $N$.
We note that the Goldstone mode is the only well-defined collective mode as it results from the spontaneous spin SU(4) symmetry breaking, though the whole spectrum we observed above exhibits a continuum. So, our EEs analysis is consistent with this wisdom. Nevertheless, for other two $q=\pi/6$ and $q=\pi/3$, the EE of the dominate modes  gradually converges to a constant up to the number of the points we calculated $N=12000$. According to the definition of the EE presented in Sec.\ref{EE}, the convergence of the EE suggests the particle-hole pair excitations are relatively free in the space of flavor $\sigma$ and $\bar{\sigma}$ respectively. So, these excitations are not confined to form magnons. As a matter of fact, the dominate modes are fractionalized excitations.


\begin{figure}
  \centering
  \includegraphics[scale=0.4]{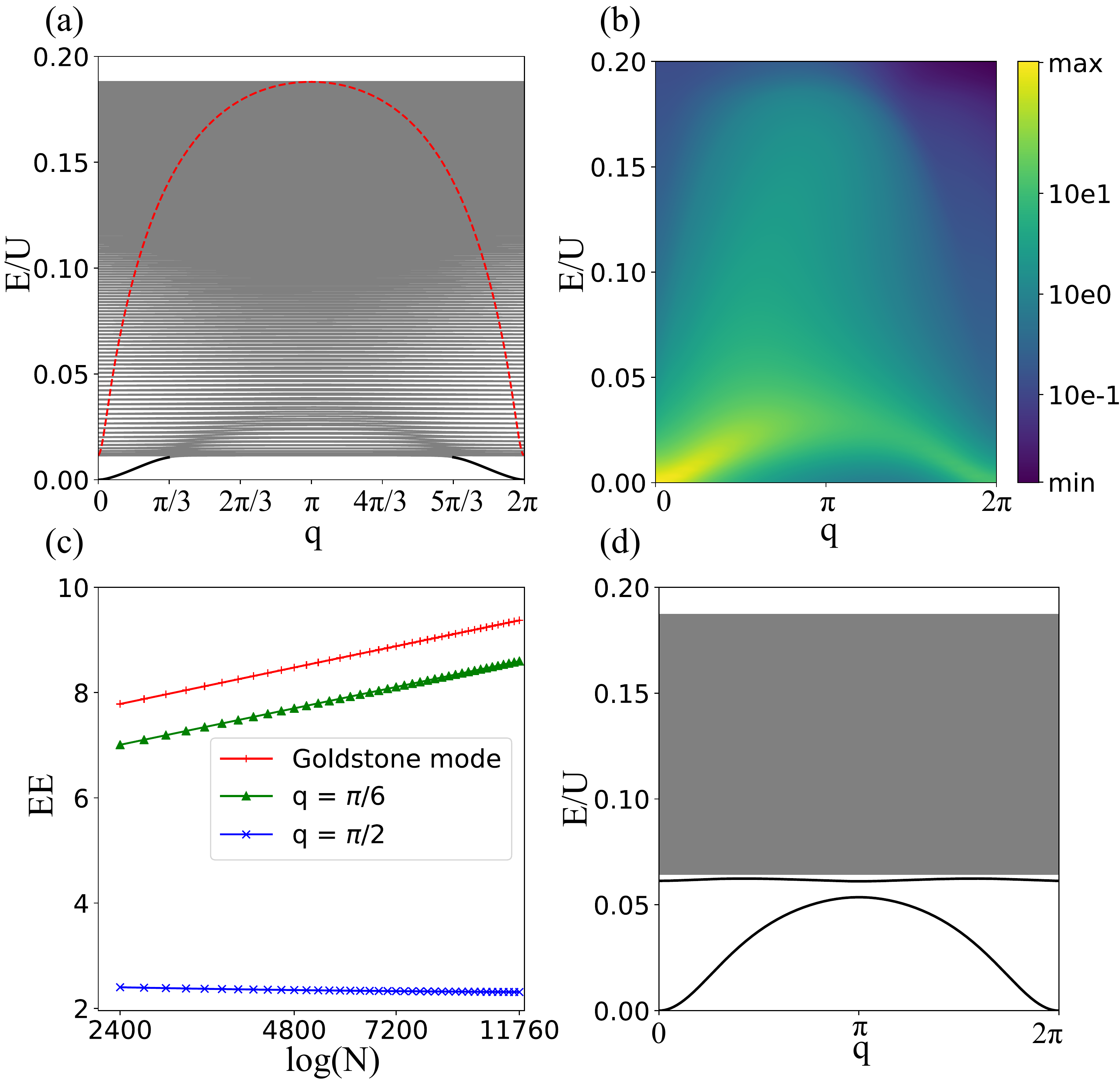}
  \caption{(a) Dispersion and (b) spectral functions of spin excitations in a ribbon system with finite width $W = 20$. (c) EE scaling behaviors of the modes at q = $\pi/2$ (blue), q = $\pi/6$ (green) and Goldstone mode q = 0 (red) with width $W=20$. (d) Spin excitation dispersion in the ribbon system with $W = 4$.}\label{fig:Fig5}
\end{figure}

\par The disappearance of the magnons is believed to come from the interactions between magnon and Stoner continuum which extends to the zero energy for the half-infinite system discussed above. If we consider a finite system, the Stoner continuum will open a gap as shown in Fig. \ref{fig:Fig5}~(a) for a system with width ($W = 20$). In this case, a single dispersion denoted by the black line can be found in the Stoner continuum gap and then merges into the continuum. From the spectral function in Fig. \ref{fig:Fig5}~(b), its spectral intensity in the gap is stronger than that in the continuum. In Fig. \ref{fig:Fig5}~(c), we plot the EE scaling behaviors of two eigenmodes, the Goldstone mode, and the mode at $q=\pi/6$, together with the dominate mode in the continuum at $q=\pi/2$ for a system with width $W=20$. While the EE for the dominate mode with $q=\pi/2$ converges to a constant, the EEs of the other two modes are logarithmically divergent with the system size. So, the eigenmode in the gap is identified as magnon. While the dominate mode, which can be traced to the magnon spoiled in the Stoner continuum [see Fig. \ref{fig:Fig5}~ (a) and (b)], are still converged.

\par When $W = 4$, the gap increases and two dispersions now appear [Fig. \ref{fig:Fig5}~(d)], which correspond to the optical and acoustic branches of magnons. For a one-dimensional edge band, the generic topological information is encoded in the Berry phase. To check their topological property, we further calculate the Berry phase of each magnon band,
\begin{equation}
\gamma=\operatorname{Im} \oint\left\langle\Psi_{q}\left|\frac{\partial}{\partial q}\right| \Psi_{q}\right\rangle dq,
\end{equation}
where $|\Psi_{q}\rangle$ denotes the corresponding eigenstate of the magnon band with momentum $q$. We find that the optical branch of the magnon band exhibits nontrivial Berry phase $\gamma = \pi$, suggesting that it has non-trivial topological properties. The existence of the topological magnon in the Stoner continuum gap shares a similarity with that in the Tasaki model \cite{PhysRevB.97.245111}.


\section{Summary and discussion}\label{SD}

\par In summary, we have studied the defect-induced edge ferromagnetism and spin excitations in the SU(4) $\pi$-flux Hubbard model on the honeycomb lattice. We have found the edge defects can induce edge flat bands that results in the SU(4) edge ferromagnetism. Furthermore, we have studied the spin excitations by projecting the Hubbard interactions onto the edge flat band. The excitations appear as a continuum covers the entire energy region and there is no well-defined magnon excitations, leaving only a broad reminiscence. Through the further EE analysis, we show that the reminiscence is the fractionalized excitations from the disappeared magnons. Moreover, in ribbon systems with finite width, the disappeared magnons can be restored in the gap opened due to the finite-size width and the optical branch of the restored magnons is shown to be topological nontrivial.
Our results about the SU(4) edge magnetism in the
thermodynamic limit lay the foundation for the spin excitations in various nanostructures, and the emergence of deconfined spinons would affect the spin-transport properties. According to our results, an ideal edge spin-transport should be based on the deconfined spinons and regulable magnons in the finite-size gap rather than the traditional magnons. Besides, the deconfined topological nontrivial magnons may also result in special spin-transport.

\section*{Acknowledgement}
\par The work was supported by National Key Research and Development Program of China (Grant
No. 2021YFA1400400), and the National Natural Science Foundation of China (Grants No.11904170, No.92165205), the Natural Science Foundation of Jiangsu Province, China (Grant No. BK20190436), and the Doctoral Program of Innovation and Entrepreneurship in Jiangsu Province.

 \bibliographystyle{elsarticle-num}
 \bibliography{ref}





\end{document}